\documentclass{article}
\usepackage{spconf,amsmath}
\usepackage[pdftex]{graphicx}
\usepackage{enumerate}
\begin{document}\sloppy

\def\x{{\mathbf x}}
\def\L{{\cal L}}

\title{Attribute-Based Multi-Dimensional Scalable Access Control For Social Media Sharing}
\name{Changsha Ma and Chang Wen Chen}
\address{Dept. of Comp. Sci. and Eng., State Univ. of New York at Buffalo, Buffalo, NY, 14260, USA\\
changsha@buffalo.edu, chencw@buffalo.edu}
\maketitle

\begin{abstract}
Media sharing is an extremely popular paradigm of social interaction in online social networks (OSNs) nowadays. The scalable media access control is essential to perform information sharing among users with various access privileges.
In this paper, we present a multi-dimensional scalable media access control (MD-SMAC) system based on the proposed scalable ciphertext policy attribute-based encryption (SCP-ABE) algorithm.
In the proposed MD-SMAC system, fine-grained access control can be performed on the media contents encoded in a multi-dimensional scalable manner based on data consumers' diverse attributes.
Through security analysis, we show that the proposed MC-SMAC system is able to resist collusion attacks. Additionally, we conduct experiments to evaluate the efficiency performance of the proposed system, especially on mobile devices.
\end{abstract}

\vspace{-2mm}
\section{Introduction}
\label{sec:intro}
Social media sharing has been an extremely popular social interaction in online social networks (OSN) nowadays. For example, on the world's most popular video sharing web site YouTube, there are more than 1 billion active users, who averagely upload 300 hours of video contents per minute \cite{youtube}. For another example, since Facebook has launched the video sharing platform in 2014, more than 1 billion videos are watched on Facebook every day \cite{facebook}.

Although most of the current social media contents are free and available for all the others, the portion of social media contents that require specific access privileges keeps increasing. The phenomena can be driven by interest or by privacy requirement. In both cases, there is a need of media access control mechanism. Specifically, since the social media contents are generally stored in the servers owned by the OSNs, while not managed by the media data distributors themselves, it is more appropriate to apply encryption-based access control instead of rule-based access control.

\begin{figure}[!t]
\centering
\includegraphics[width=2.6in]{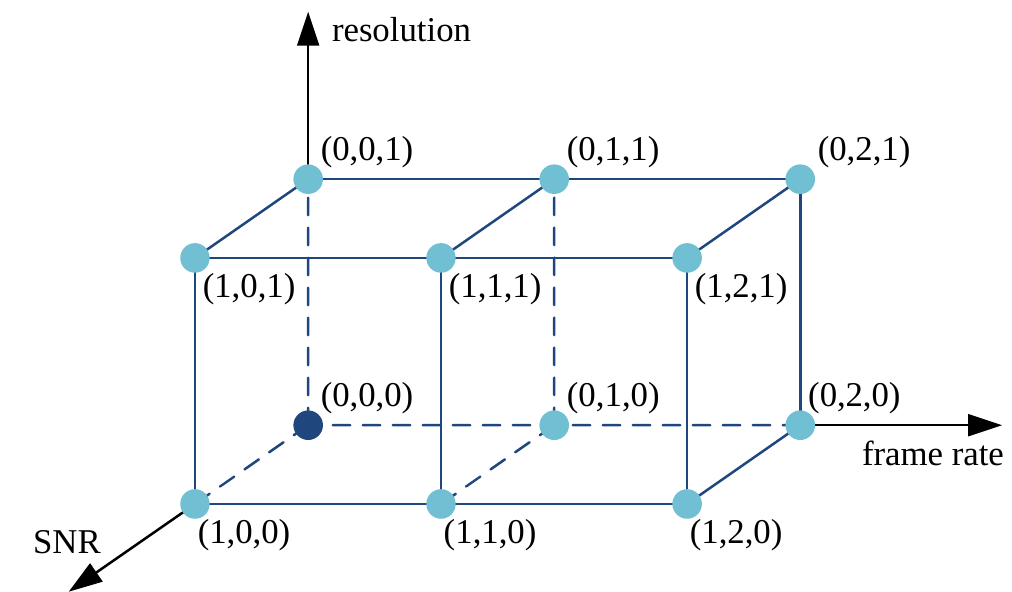}
\caption{2-by-3-by-2 scalable media data structure}
\label{intro}
\vspace{-2mm}
\end{figure}

Additionally, due to the diversity of social attributes and the heterogeneity of network environment associated with different social network users, it is reasonably desired to distribute a media content in different versions with different qualities \cite{multid}. With the support of scalable media formats such as JPEG 2000 and H.264 scalable video coding (SVC), the media stream can be encoded into a base layer that provides the lowest quality, and enhancement layers for enhancing the media data quality in terms of multiple factors such as frame rate, resolution, and SNR \cite{svc}. Take the 2-by-3-by-2 data structure shown in Fig.\ref{intro} as an example, the base layer (0,0,0) provides the media quality with the lowest SNR, the lowest frame rate, and the lowest resolution. With an additional enhancement layer (1,0,0) ((0,1,0), (0,0,1)), the media quality improves in terms of SNR, (frame rate, resolution). More enhancement layers corresponds to higher media qualities. We hence need to assign an access key to each layer, and set an access policy for each media quality. Therefore, under the scalable media data structure, the access privileges are no longer just presence and absence, but instead scalable in multiple dimensions and with the total number of $2\times 3 \times 2$ in this example. The multi-dimensional scalable media access control (MD-SMAC), which is able to enforce the multiple-dimensional scalable access policies, is therefore expected.
In this paper, we propose a MD-SMAC system based on the proposed scalable ciphtertext policy attribute-based encryption (SCP-ABE) algorithm, which encrypts multiple messages under a scalable attribute-based access tree. Specifically, we assign an access key for each layer of the scalably encoded media content, and then encrypt the key under SCP-ABE. The access on a media layer thus requires decrypting the corresponding access key under SCP-ABE. A successful decryption requires the attributes described in the corresponding part of the SCP-ABE access tree. The more parts an individual's attributes can satisfy the access tree, the more media layers the individual can decrypt, and hence the higher media quality the individual can enjoy.

The rest of the paper is organized as follows. In Section 2 we introduce the related work. In Section 3 we present the proposed MD-SMAC system for social media sharing. Security and performance analysis are presented in Section 4. Finally, conclusions are summarized in Section 5.
\vspace{-2mm}
\section{Related Work}
\vspace{-2mm}
As we know, when a media content is encoded in a scalable manner, there will be multiple layers generated. As a consequence, key distribution will become troublesome since multiple access keys may be sent to one data consumer. In order to reduce the cost of key distribution, related works \cite{multid, dhscalalbe, scalable1, scalable3} usually expand one access key as multiple segments. The number of segments is equal to the number of dimensions. Additionally, segments of lower-level access keys can be generated from the corresponding segments of higher-level access keys through one-way hash chains. In this way, only one access key needs to be sent to each data consumer. Hence, the key distribution cost suppose to significantly decrease with slightly sacrificing the computation cost.

Unfortunately, such method may result in user collusion when the media layers are organized following the multi-dimensional scalable format. That is two users may collude with each other to generate a valid but illegal access key for unauthorized higher level access. The resistance of such collusion attack requires increasing the number of key segments. From the results shown in \cite{multid}, the best performance of a collusion secure scheme needs $mn$ key segments under a $m$-by-$n$-by-$k$ ($k = max(m,n,k)$) data structure, but at the same time results in $O(mnk)$ more hash calculations. As a result, the communication cost saved by segmenting access keys is less significant, while the resulted computation cost is more than expected. Therefore, the aim to reduce the cost of key distribution is not sufficiently meaningful.

We hence consider a more important issue of scalable access control under the social media sharing application as how to distribute the access keys among media data consumers. To solve this problem, a recent research \cite{mcpabe} has reported a multi-message ciphertext policy attribute-based encryption (MCP-ABE) scheme to achieve fine-grained scalable media access control. Specifically, this scheme sets a scalable access policy for the scalable media data, and encrypts the access keys under the policy. Data consumers with proper attributes can obtain the specific access keys and then acquire the media data with the corresponding quality. This scheme is rather flexible in terms of key distribution, since the data distributor does not need to distribute access keys for each consumer. However, this scheme only considers the media data with one-dimensional scalability. In \cite{2dcpabe}, a 2-dimensional scalable access control scheme is proposed by converting the 2-dimensional scalable data structure into a tree structure. But there is no discuss on how to expand the scheme to achieve MD-SMAC.  This issue is addressed in our work. More specifically, we propose a MD-SMAC system based on the proposed SCP-ABE algorithm, which is able to adapt to data structures that are scalable in terms of arbitrary dimensions.
\vspace{-2mm}
\section{The Proposed MD-SMAC System}
\begin{figure*}[!t]
\centering
\includegraphics[width=6in]{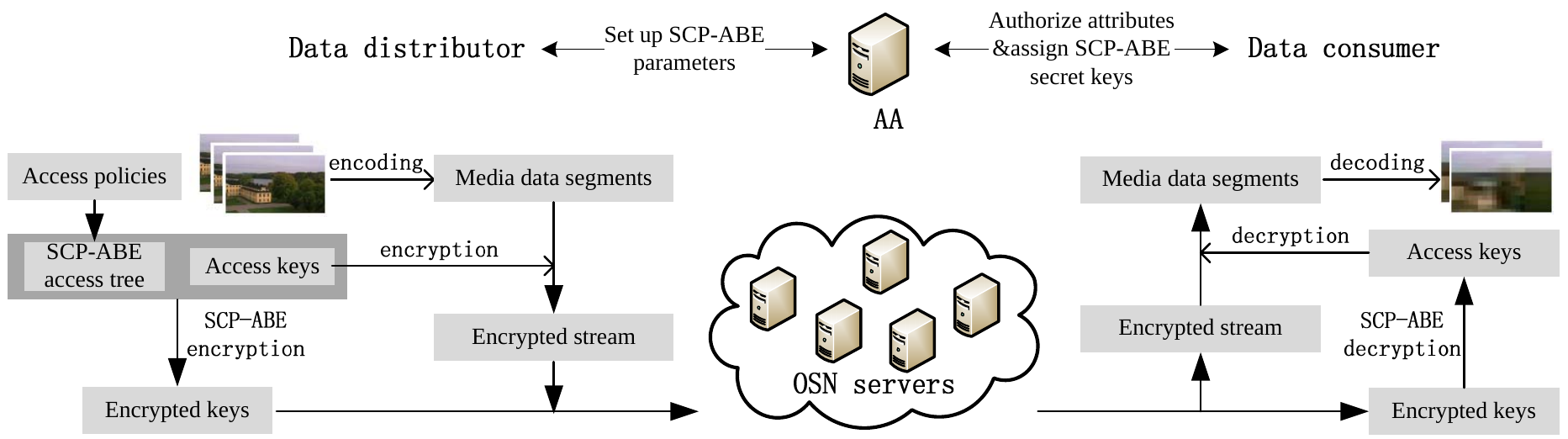}
\caption{Multi-Dimensional Scalable Social Media Sharing And Access Control}
\label{architecture}
\end{figure*}
\subsection{System Overview}
\vspace{-1mm}
In the proposed scalable media access control architecture as shown in Fig.\ref{architecture}, there are four communication parties, i.e. the data distributor who owns the copyright of the media contents, the data consumers who are interested in the media contents, the OSN servers run by the social networking site, and the attribute authority (AA) who authorizes the attributes involved in the access policies.
In practice, the AA can be run by a trusted third party. Or alternatively, data distributors can run their AAs by themselves. In either way, the AA should be guaranteed accessible by the data consumers.

In the system, a media stream is encoded into a manner that the quality is scalable in multiple dimensions. Each resulted media layer is encrypted by an access key\footnote{The encryption algorithm is out of the scope of our discussion.}, which is further encrypted under the SCP-ABE algorithm. The encrypted media stream and the encrypted access keys are stored on the OSN servers, and accessible for all data consumers. A data consumer decrypts the media stream by first decrypting the access keys using their SCP-ABE user keys that are assigned by the AA according to their attributes.
\vspace{-2mm}
\subsection{Access Privileges in The System}
The principle of scalable media access control is that the access to the higher level media layer guarantees the access to the lower level media layer \cite{multid}.
In attribute-based access control schemes, attributes determine access privileges. Therefore, to achieve scalable access control, the attributes associated with the higher level layer should also satisfy the access policy to the lower level layer. In a 3D-scalable media data structure, e.g. the 2-by-3-by-2 one shown in Fig.\ref{intro}, suppose the access policy associated with the layer $ijk (i = 0,1; j = 0,1,2; k = 0,1)$ is $P_{ijk}$, then we always have:
\begin{equation}
P_{ijk} \subseteq P_{pql}, i\leq p, j \leq q, k \leq l\nonumber
\end{equation}
Based on this rule, we can further conclude that:
\begin{equation}
 P_{ijk} \subseteq P_{pql} \cap P_{mng}, i\leq min(p,m), j \leq min(q,n), k \leq min(l,g)\nonumber
\end{equation}
 We follow these access privilege enforcement principles when designing the SCP-ABE algorithm. For the 2D-scalable media data structure shown in Fig.\ref{tree}, we hence have $P_{11} (I_1) \subseteq P_{12}\cap P_{21} (I_2)$, $P_{12} \subseteq P_{13}$, and $P_{21} \subseteq P_{22}$. Furthermore, we have $I_2 \subseteq P_{13}\cap P_{22} (I_3)$. Similarly, we have $I_3 \subseteq I_4$.
\subsection{The SCP-ABE algorithm}
The proposed SCP-ABE algorithm is a variation of the CP-ABE algorithm \cite{cpabe}. In CP-ABE scheme, a message is encrypted under an access policy and a secret key of a user is associated with a set of attributes, which are authorized by attribute authorities (AAs). A user could decrypt the message only if his or her attributes satisfy the access policy.
Instead of encrypting one message in one ciphertext as in CP-ABE, SCP-ABE encrypts multiple ones in one ciphertext. Under the proposed MD-SMAC system, a message refers to an access key that is used to encrypt a media layer.
The messages a user can decrypt depends on the individual's access privilege, which is determined by the individual's attributes.

Similar to CP-ABE, SCP-ABE algorithm is also based on the bilinearity property and the non-degeneracy property of bilinear map. Moreover, SCP-ABE algorithm is composed of six sub-algorithms including system setup, access tree construction, encryption, decryption, user key generation, and delegation. We present SCP-ABE by first introducing the properties of bilinear map, and then describing each sub-algorithm.
\vspace{-2mm}
\subsubsection{Bilinear Map}

Let $G_{0}$ and $G_{1}$ be two multiplicative cyclic groups of prime order $p$. Let $g$ be a generator of $G_{0}$ and $e$ be a bilinear map, $e: G_{0}\times G_{0} \rightarrow G_{1}$. Then $e$ has the following properties:
\vspace{-1mm}
\begin{itemize}
  \item \emph{Bilinearity}: for all $u, v \in G_{0}$ and $a, b \in Z_{p}$, we have $e(u^{a}, v^{b})=e(u,v)^{ab}$.
  \item \emph{Non-degeneracy}: $e(g, g)\neq 1$.
\end{itemize}
\vspace{-2mm}
\subsubsection{System Setup}
\vspace{-2mm}
The setup algorithm chooses a bilinear group $G_{0}$ of prime order $p$ with generator $g$, and two random exponents $\alpha, \beta \in Z_{p}$. The public key $PK$ and the master key $MK$ are returned as: $\{PK=G_{0}, g, h=g^{\beta}, f=g^{1/\beta}, e(g,g)^{\alpha}\}$, $\{MK=\beta, g^{\alpha}\}$.
\vspace{-2mm}
\subsubsection{Access Tree Construction}
\begin{figure*}
\centering
\includegraphics[width=6.7in]{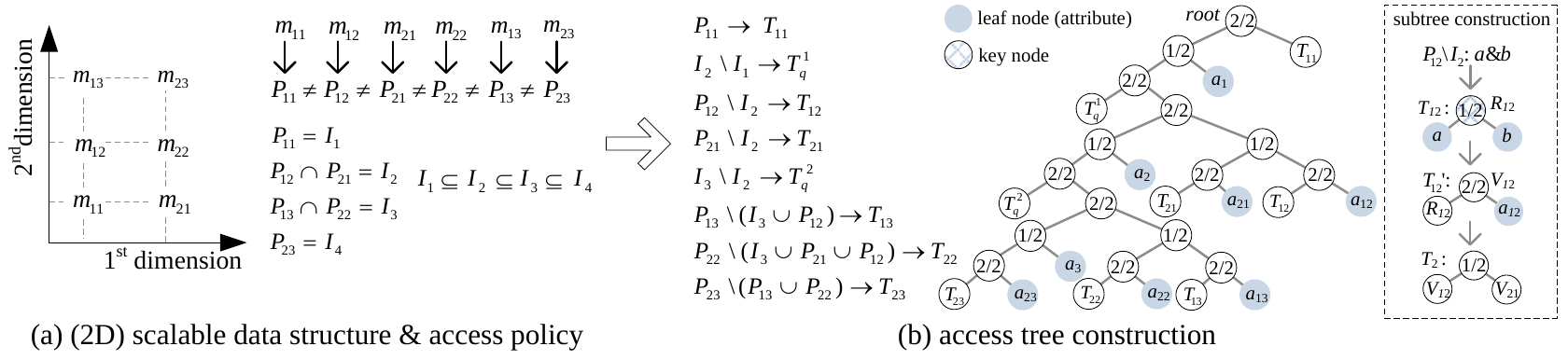}
\caption{An indication of SCP-ABE access tree construction}
\label{tree}
\end{figure*}
\vspace{-2mm}
Given the scalable data structure and the corresponding access policies, the access key generation algorithm constructs the access tree, which has the feature that the access to higher level data layer guarantees the access to lower level data layer, but not vice versa. For ease of presentation, we here use the 2D-scalable ($M$-by-$N$) data structure to describe the access tree construction algorithm. However, note that the process is applicable for multiple-dimensional scalable data structure.

Suppose that the access policy corresponding to each data segment $m_{ij}$ $(i = 1, 2, ..., M; j = 1, 2, ..., N)$ is $P_{ij}$. The access policy has the property that $P_{ij} < P_{mn}$ if $i+j < p+q$, meaning that all rules in $P_{ij}$ are included in $P_{pq}$ if $m_{pq}$ has longer distance to $m_{11}$ than $m_{ij}$. Then the access tree $T$ is constructed as following.
\begin{enumerate}[a.]
\item To begin with, we define the \emph{referee} for $m_{ij} (i = 1, 2, ..., M; j = 1, 2, ..., N)$ as the key(s) in the path from $m_{ij}$ to $m_{11}$ that is (are) nearest to $m_{ij}$. The referee for $m_{11}$ is defined as itself. In the example shown in Fig. 5, the referees of $m_{12}, m_{21}, m_{13}, m_{22}$, and $m_{23}$ are ($m_{11}$), ($m_{11}$), ($m_{12}$), ($m_{12}, m_{21}$), and ($m_{13}, m_{22}$), respectively. Additionally, we divide the data segments into $M+N-1$ groups according to their distances to $m_{11}$ in the data structure. Under the data structure shown in Fig. 5, there are four groups, i.e. $G_1=(m_{11})$, $G_2=(m_{12}, m_{21})$, $G_3=(m_{13}, m_{22})$, and $G_4=(m_{23})$. Let $U_i (i = 1, ..., M+N-1)$ be the union access policy of members in group $G_i $, and $I_i$ be the common access policy of members in group $G_i$.
\item Construct the sub-tree $T_q^i (i = 1, 2, ..., M+N-3)$ according to $I_{i+1} \setminus I_i$.
\item Then we construct $T_{11}$ according to $P_{11}$, and $T_{ij} (i = 2, ..., M | j=1; j = 2, ..., N | i=1) $ according to $P_{ij}\setminus (I_k \cup P_{ij}')$, where $I_k$ is the common access policy of members in group $G_k$ that $P_{ij} $ belongs to, and $P_{ij}'$ is the referee(s) of $P_{ij}$. The root $R_{ij} (i = 1, 2, ..., M; j = 1, 2, ..., N)$ of $T_{ij}$ is called key node.
\item Let $R_{ij} (i = 2, ..., M | j=1; j = 2, ..., N | i=1)$ and an additional attribute $a_{ij}$ be the children of the $and$ gate, build the subtree $T_{ij}'$, whose root is $V_{ij}$.
\item Let all $V_{ij}$ in the group $G_i (i = 2, 3, ..., M+N-2)$ be the children of the $1/|G_i|$ gate, build the group subtree $T_{i}$. Let $T_{M+N-1}$ be $T_{MN}'$, and $T_1$ be $T_{11}$.
\item The whole access tree is constructed group by group in a bottom-to-up manner. Starting from $i=M+N-2$, for group $G_i$, let an $and$ gate be the mother of the root of $T_{i}$ and an $or$ gate whose children are a selected attribute $a_{i}$ and the root of $T_{i+1}$. Then let the $and$ gate and $T_q^{i-1}$ be the children of a new $and$ gate. Let the new constructed tree as $T_i$, and recursively do the previous steps until all groups are in the access tree. The root of the access tree will be an $and$ gate.
\end{enumerate}

Then we choose a polynomial $p_{x}$ for each tree node $x$ in the access tree $\mathcal{T}$ in a manner as: (1) From the bottom to up, set the degree $d_{x}$ of the polynomial $p_{x}$ to be $d_{x}=k_{x}-1$, where $k_{x}$ is the threshold value of node $x$; (2) For the root node $root$, choose a random $s\in Z_{p}$ and set $p_{root}(0)=s$, and randomly choose other points of polynomial $p_{root}$; (3) For any other node $x$, set $p_{x}(0)=p_{parent(x)}(index(x))$, and randomly choose other points of $p_{x}$.
Figure \ref{tree} indicates a case of access tree construction based on a 2-by-3 data structure.

In the access tree structure, we have added some additional attributes except for the attributes related to the access policies. Specifically, $a_i (i=1,2,...,M+N-2)$ is used to enable users to compute starting from the level where $G_i$ locates. For example, if the attributes of a user only conform to $P_{11}$, the user takes the user key corresponding to attributes in $P_{11}$ and $a_1$ as input of the decryption algorithm, and performs decryption operation starting from the level that $G_1$ locates. If the user has attributes that conform to $P_{12}$, he or she can instead perform decryption starting from the level that $G_2$ locates. Additionally, $a_{ij} (i = 2, ..., M | j=1; j = 2, ..., N | i=1)$ is used to guarantee the uniqueness of each key node, which is essential to enforce the access policies. For example, without $a_{12}$ and $a_{21}$, $p_{R_{12}}(0)$ will be equal to $p_{R_{21}}(0)$ since their mother node has the degree of zero. As a result, a user who can access $p_{R_{12}}(0)$ will also be able to access $p_{R_{21}}(0)$ even without the required attributes.
\vspace{-2mm}
\subsubsection{Encryption}
\vspace{-2mm}
The encryption algorithm encrypts the data segments. Let $L$ be the set of leaf nodes in $\mathcal{T}$, $K$ be the set of key nodes $V_{ij}$ $(i = 1, 2, ..., M; j = 1, 2, ..., N)$, and $m_{ij}$ be the corresponding data segment, the ciphertext is given as
\begin{equation}
\begin{split}
&CT = (\mathcal{T}, \forall i\in L: E_{i}=g^{p_{i}(0)}, E_{i}'=H(att(i))^{p_{i}(0)}\\
&\forall R_{ij}\in K: \tilde{C_{ij}}=m_{ij} e(g,g)^{\alpha (p_{R_{ij}}(0)+s)}, C_{ij}=h^{p_{R_{ij}}(0)+s})\nonumber
\end{split}
\end{equation}
\vspace{-2mm}
\subsubsection{Decryption}
\vspace{-2mm}
The decryption algorithm takes as input several messages encrypted under an access policy, a secret key $SK$ of a user, and the public key $PK$.
The number of access keys that a user is able to decrypt depends on how much the attributes can satisfy the access policy.
Suppose that the highest level of access keys a user can obtain is $m_{pq}$ $(1\leq p \leq M, 1\leq q \leq N)$. Starting from $T_{pq}$, the user needs to perform the following computation for each leaf node $x$ of $T_{pq}$ and that locates at the path from $R_{pq}$ to the root.
\begin{equation}
\vspace{-1mm}
\begin{split}
F_{x}& = \frac {e(D_{x}, E_{x})} {e(D_{x}', E_{x}') }\\
& = \frac {e(g^{r}\cdot H(attri_{x})^{r_{x}}, g^{p_{x}(0)})}{e(g^{r_{x}}, H(attri_{x})^{p_{x}(0)})}\\
& = \frac {e(g^{r},g^{p_{x}(0)}) \cdot e(H(attri_{x})^{r_{x}}, g^{p_{x}(0)})}{e(g^{r_{x}}, H(attri_{x})^{p_{x}(0)})}\\
&= e(g,g)^{rp_{x}(0)}\nonumber
\end{split}
\end{equation}
Then the user recursively computes the corresponding values of non-leaf nodes including the key nodes, in a bottom-up manner using polynomial interpolation technique \cite{cpabe}, until reaches the root and obtains:
\begin{equation}
F_{root} = e(g,g)^{rp_{root(0)}} = e(g,g)^{rs}\nonumber
\end{equation}
In this process, the user also obtains:
\begin{equation}
F_{R_{ij}} = e(g,g)^{rp_{R_{ij}}(0)} (i = 1, 2,..., p, j = 1, 2,..., q)\nonumber
\end{equation}
Furthermore, the user compute $ K_{ij}= F_{R_{ij}} \cdot F_{root} = e(g,g)^{r(p_{R_{ij}}(0)+s)}$. Each data segment $m_{ij}$ $(i = 1, 2,..., p, j = 1, 2,..., q)$ can then be decrypted as following.
\begin{equation}
\vspace{-2mm}
\begin{split}
  &\ \ \ \ \ \frac{\tilde{C_{ij}}}{e(C_{ij},D)/K_{ij}} \\
   &= \frac{m_{ij} e(g,g)^{\alpha (p_{R_{ij}}(0)+s)}}{e(h^{p_{R_{ij}}(0)+s},g^{(\alpha+r)/\beta})/e(g,g)^{r(p_{R_{ij}}(0)+s)}}\\
   &= \frac{m_{ij} e(g,g)^{(\alpha+r) (p_{R_{ij}}(0)+s)}}{e(g^{\beta (p_{R_{ij}}(0)+s)},g^{(\alpha+r)/\beta})}\\
   &= \frac{m_{ij} e(g,g)^{(\alpha+r) (p_{R_{ij}}(0)+s)}}{e(g,g)^{\beta (p_{R_{ij}}(0)+s)\cdot(\alpha+r)/\beta}}\\
   & = m_{ij}\nonumber
\end{split}
\end{equation}

\subsubsection{User Key Generation}
The algorithm generates user secret keys used for deriving the access keys. Specifically, taking a set of attributes $S$ as input, the key generation algorithm outputs a secret key that identifies with the set. The algorithm selects a random $r_{i} \in Z_{p}$ for every attribute in $S$. Note that $S$ always include $a_i (i=1,2,...,M+N-2)$, and $a_{ij} (i = 2, ..., M | j=1; j = 2, ..., N | i=1)$. The user secret key is computed as
\begin{equation}
\begin{split}
SK = \{&D = g^{(\alpha + r)/\beta}, \\
&\forall i \in S: D_{i}=g^{r}\cdot H(attri_i)^{r_{i}}, D_{i}'=g^{r_{i}}\}\nonumber
\end{split}
\vspace{-2mm}
\end{equation}
\subsubsection{Delegation}
Given a user secret key $SK$ with the attribute set $S$, the delegation algorithm creates a new user secret key $\tilde{SK}$ with the attribute set $\tilde{S} \subseteq S$. Specifically, the algorithm selects a random number  $\tilde{r} \in Z_{p}$ and also $\tilde{r_{i}} \in Z_{p} \forall i \in \tilde{S}$. Then $\tilde{SK}$ is created as
\begin{equation}
\begin{split}
\tilde{SK} = \{&\tilde{D} = Df^{\tilde{r}}, \\
& \forall i \in \tilde{S}: \tilde{D_{i}}=D_{i}g^{\tilde{r}}\cdot H(attri_i)^{\tilde{r_{i}}}, \tilde{D_{i}}'=D_{i}'g^{\tilde{r_{i}}}\}\nonumber
\end{split}
\vspace{-2mm}
\end{equation}

\subsection{Discussion}
\vspace{-2mm}
Following the access privilege enforcement principle, the attributes associated with a higher level access key always contains those associated with a lower level access key. If we simply apply an existing ABE algorithm to encrypt each access key separately, the computation power will be significantly wasted on repeatedly computing on the overlapped attributes. Such redundant computations can be avoided in the proposed SCP-ABE algorithm, since the access tree is organized in a scalable manner. Therefore, the proposed SCP-ABE algorithm is essential for the efficiency of the MD-SMAC system.

Additionally, the delegation algorithm in SCP-ABE provides additional flexibilities for social media sharing. For example, if a media data distributor individually runs the AA, which is not reachable for friends' friends, the distributor's friends can run the delegation algorithm and assign user keys for their friends. The detailed discussion will be presented in our future work.
\vspace{-2mm}
\section{Performance Evaluation}
\subsection{Security Performance}
\vspace{-2mm}
By encrypting the media layers resulted from scalable media format under SCP-ABE, the proposed MD-SMAC system enables a data distributor to define fine-grained and scalable access privileges on the media contents shared on the OSNs. In this process, there is a potential collusion issue that two or more data consumers may collude with each other and try to obtain an access privilege higher than they suppose to deserve. In this section, we present analysis on how collusion resistance is provided by the proposed MD-SMAC system. Specifically, we consider collusion in the following two cases. For ease of presentation, we use the media data structure shown in Fig. \ref{tree}.
\vspace{-3mm}
\subsubsection{Cross Group Collusion}
\vspace{-1mm}
A classical collusion issue in SMAC is that the user who has access key $m_{ij}$ (e.g. $m_{13}$) and other lower level access keys (e.g. $m_{11}$ and $m_{12}$) collude with the user who owns access key $m_{pq}$ (e.g. $m_{21}$) and other lower level access keys (e.g. $m_{11}$), trying to derive the target access key $m_{kl}$ ($k = max(i,p), l = max(j,q)$, e.g. $m_{23}$) and all lower level access keys that the two users do not have (e.g. $m_{22}$).
Since the target access key always locate at a different group (as defined in SCP-ABE access tree construction algorithm), we refer such collusion as cross group collusion.

Since the access keys are encrypted under SCP-ABE, the success of such collusion requires the attributes of the two users satisfy the access policy corresponding to the target access key. For example, the derivation of $m_{23}$ from $m_{13}$ and $m_{21}$ needs $P_{23} \subseteq P_{12} \cup P_{21}$. Under the access privilege enforcement principle, we have $P_{12} \cup P_{21} \subseteq P_{22} \subseteq P_{23}$. The success of collusion then requires $P_{23} = P_{12} \cup P_{21}$, which means there is no $T_{22}'$ or $T_{23}'$ in the access tree and hence no $K_{22}$ or $K_{23}$. However, to avoid that $m_{22}$ and $m_{23}$ are transmitted in plain text, they have to be encrypted under SCP-ABE. Therefore, such collusion is infeasible.
\vspace{-2mm}
\subsubsection{Random Collusion}
\vspace{-1mm}
The other potential collusion issue is that two users exchange their attributes to obtain a specific access privilege. For example, suppose $P_{11}$ is $b_0$ \emph{and} $b_1$. One user owns attributes $b_1$, while the other user just owns the attribute $b_0$. The two users then try to collude and decrypt $m_{11}$. For the first user, he has the following secret key and is able to compute $F_{b_1}=e(g,g)^{r_ap_{b_1}(0)}$ for leaf node $b_1$ in $T_{11}$.
\begin{equation}
\begin{split}
SK_a = \{&D_a = g^{(\alpha + r_a)/\beta}, \\
&D_{1}=g^{r_a}\cdot H(b_1)^{r_{1}}, D_{1}'=g^{r_{1}}\}\nonumber
\end{split}
\end{equation}

\noindent For the second user, he has the following secret key and is able to compute $F_{b_0}=e(g,g)^{r_bp_{b_0}(0)}$ for leaf node $b_0$ in $T_{11}$.
\begin{equation}
\begin{split}
SK_b = \{&D_b = g^{(\alpha + r_b)/\beta}, \\
&D_{2}=g^{r_b}\cdot H(b_0)^{r_{2}}, D_{2}'=g^{r_{2}}\}\nonumber
\end{split}
\end{equation}
Their goal is to obtain $F_{R_{11}}$, which is essential to derive $K_{11}$ and to decrypt $m_{11}$.  However, $F_{R_{11}}$ should be either $e(g,g)^{r_ap_{R_{11}}(0)}$ or $e(g,g)^{r_bp_{R_{11}}(0)}$. If the first user wants to compute $F_{R_{11}}$, he needs the second user to send him $e(g,g)^{r_ap_{b_0}(0)}$, instead of $e(g,g)^{r_bp_{b_0}(0)}$ that the second user has. Therefore, the first user cannot obtain $F_{R_{11}}$. Similarly, the second user cannot obtain $F_{R_{11}}$ either.

Therefore, the random collusion is also infeasible in our proposed system.
\vspace{-3.5mm}
\subsection{Efficiency Performance}
\vspace{-1mm}
In the proposed MD-SMAC system, a media data distributor always first encrypts their media content before distribution, while a media data consumer always needs to perform decryption before enjoying the media content. The encryption and decryption operations on online media contents actually have already been widely applied in web sites nowadays with the emergence of security requirement. Therefore, the extra cost in the proposed MD-SMAC system merely results from the operations under SCP-ABE. In this section, we present analysis and experimental results of the SCP-ABE computation cost on both data distributor side and data consumer side.
\vspace{-7mm}
\subsubsection{Distributor Side Computation Cost}
\vspace{-2mm}
Distributor side computation cost mainly comes from SCP-ABE encryption, which involves pairings, exponentiations, and multiplications computations on $G_0$ and $G_1$. The encryption computation cost linearly increases with the number of attributes.
We measure the average computation time per attribute on the user side on Google Nexus 4 (1.5GHz quad-core Snapdragon S4 Pro with Krait CPUs, Android 5.0), and on Lenovo T430 (2.6GHz Intel i5 CPU, Ubuntu 14.04)\footnote{As in \cite{cpabe}, operations are conducted using a 160-bit elliptic curve group based on the curve $y^2 = x^3 + x$ over 512-bit finite field.}, which are 60ms and 35ms respectively.
Suppose the given media content has $M$ layers, and the access policy of layer $i$ involves $n_i$ ($n_{i+1} > n_i$) attributes. Plus the additional $N$ attributes involved in the SCP-ABE algorithm, there will be $n_M + N$ attributes in total. The total computation time of SCP-ABE encryption and decryption can be obtained accordingly.
\vspace{-3mm}
\subsubsection{Consumer Side Computation Cost}
\vspace{-2mm}
Consumer side computation cost results from SCP-ABE decryption, which include pairings and multiplications computations on $G_1$. For SCP-ABE decryption, the computation cost is not only impacted by the number of attributes, but also by the access policies. For simplification, we adjust the access policies and make the computation cost grow with the number of attributes in approximately linear way, and test the computation time per ten attributes on devices. The result is 48ms on smart phone and 25ms on laptop.
The server side computation cost comes from delegation. The AA side computation cost comes from user key generation. Both of the two algorithms involve in exponentiations and multiplications on $G_0$ and $G_1$. According to the results provided in \cite{cpabe}, the computation time of user key generation is about 30ms per attribute on workstation with 64-bit 3.2 Ghz Pentium 4 processor. The computation time of delegation is almost the same.
\vspace{-3mm}
\section{Conclusion}
\vspace{-3mm}
We have presented in this paper a novel scalable ciphertext policy attribute-based encryption (SCP-ABE) algorithm, and based on what designed a multi-dimensional scalable media access control (MD-SMAC) system. With this system, we are able to achieve secure and fine-grained access control on social media contents encoded in multi-dimensional scalable format. We have confirmed the effectiveness of the proposed system through both numerical analysis and mobile terminal implementation with typical laptop and smart phone.


\vspace{-4mm}
\begin{thebibliography}{10}
\vspace{-2.8mm}
\bibitem{youtube}
http://expandedramblings.com/index.php/youtube-statistics/
\vspace{-2.8mm}
\bibitem{facebook}
http://www.businessinsider.com/facebook-video-statistics-2015-1
\vspace{-2.8mm}
\bibitem{multid}
Zhu, X.; Chen, C. W., ``A collusion resilient key management scheme for multi-dimensional scalable media access control," Image Processing (ICIP), the 18th IEEE Int. Conference on, pp.2769,2772, 11-14 Sept. 2011.
\vspace{-2.8mm}
\bibitem{svc}
Schwarz, H.; Marpe, D.; Wiegand, T., ``Overview of the Scalable Video Coding Extension of the H.264/AVC Standard," Circuits and Systems for Video Technology, IEEE Trans. on, vol.17, no.9, pp.1103,1120, Sept. 2007.
\bibitem{dhscalalbe}
\vspace{-2.8mm}
Zhu, B.B.; Feng, M.; Li, S., ``An efficient key scheme for layered access control of MPEG-4 FGS video," ICME '04. 2004 IEEE International Conference on, pp.443,446, 30-30 June 2004.
\vspace{-2.8mm}
\bibitem{scalable1}
Imaizumi, S.; Fujiyoshi, M.; Abe, Y.; Kiya, H., ``Collusion Attack-Resilient Hierarchical Encryption of JPEG 2000 Codestreams with Scalable Access Control," Image Processing, 2007. ICIP 2007. IEEE International Conference on , vol.2, no., pp.II - 137,II - 140, Sept. 16 2007-Oct. 19 2007.
\vspace{-2.8mm}
\bibitem{scalable3}
Crampton, J.; Daud, R.; Martin, K. M., ``Constructing Key Allignment Schemes from Chain Partitions", Proc. the 24th annual IFIP WG 11.3 working conference on Data and applications security and privacy, 2010.
\vspace{-2.8mm}
\bibitem{mcpabe}
Wu, Y.; Zhuo, W.; Deng, R., ``Attribute-Based Access to Scalable Media in Cloud-Assisted Content Sharing Networks," Multimedia, IEEE Transactions on , vol.15, no.4, pp.778,788, June 2013.
\vspace{-3mm}
\bibitem{2dcpabe}
Ma, C.; Chen, C. W., ``Secure media sharing in the cloud: Two-dimensional-scalable access control and comprehensive key management," Multimedia and Expo (ICME), 2014 IEEE International Conference on, pp.1-6, 14-18 July 2014.
\vspace{-2.8mm}
\bibitem{cpabe}
Bethencourt, J.; Sahai, A.; Waters, B., ``Ciphertext-Policy Attribute-Based Encryption," Security and Privacy, 2007. SP '07. IEEE Symposium on, pp.321,334, 20-23 May 2007.
\end{thebibliography}
\end{document}